\documentclass{emulateapj}
\usepackage{graphicx}

\begin{document}

\def\ba{\begin{eqnarray}}
\def\ea{\end{eqnarray}}
\def\etal{et al.\ \rm}
\def\tpar{\tau_\parallel}

\title{Global modeling of radiatively driven accretion of metals 
from compact debris disks onto the white dwarfs.}

\author{Konstantin V. Bochkarev\altaffilmark{1}}
\altaffiltext{1}{Departmen of General and Applied Physics, Moscow 
Institute of Physics and Technology, Dolgoprudny, 141700, Russia; 
bochkarevkv@gmail.com}
\author{Roman R. Rafikov\altaffilmark{2,3}}
\altaffiltext{2}{Department of Astrophysical Sciences, 
Princeton University, Ivy Lane, Princeton, NJ 08540; 
rrr@astro.princeton.edu}
\altaffiltext{3}{Sloan Fellow}


\begin{abstract}
Recent infrared observations have revealed presence of compact (radii
$\lesssim R_\odot$) debris disks around more than a dozen of metal-rich 
white-dwarfs (WD), likely produced by tidal disruption of asteroids. 
Accretion of high-Z material from these disks may account for the metal 
contamination of these WDs. It was previously shown using local 
calculations that the Poynting-Robertson (PR) drag acting on the 
dense, optically thick disk naturally drives metal accretion 
onto the WD at the typical rate $\dot M_{PR} \approx 10^8$ g s$^{-1}$. 
Here we extend this local analysis by exploring global evolution 
of the debris disk under the action of the PR drag
for a variety of assumptions about the disk properties. We find that 
massive disks (mass $\gtrsim 10^{20}$ g), which are optically thick to  
incident stellar radiation inevitably give rise to metal accretion at rates 
$\dot M \gtrsim 0.2\dot M_{PR}$. The magnitude of $\dot M$ and its 
time evolution are determined predominantly by the initial pattern of 
the radial distribution of the debris (i.e. ring-like vs. disk-like) 
but not by 
the total mass of the disk. The latter determines only the disk 
lifetime, which can be several Myr or longer. Evolution of an optically 
thick disk generically results in the development of a sharp outer edge 
of the disk. We also find that the low mass ($\lesssim 10^{20}$ g), 
optically thin disks exhibit $\dot M\ll \dot M_{PR}$ and evolve on 
characteristic timescale $\sim 10^5-10^6$ yr, independent of their 
total mass.
\end{abstract}

\keywords{White dwarfs --- Accretion, accretion disks --- Protoplanetary disks}


\section{Introduction.}  
\label{sect:intro}


Recent ground based and {\it Spitzer} observations 
(Zuckerman \& Becklin 1987; Graham \etal 1990; Farihi \etal 2010)
have revealed the presence of infrared excesses in spectra of a 
number of metal-rich white dwarfs (WD). These excesses have
been interpreted (Jura 2003) as arising from reprocessing of  
stellar emission by a compact disk of refractory debris orbiting 
the WD. Detailed spectral modeling (Jura 2003; Farihi \etal 2010) suggests
that these disks are geometrically thin and essentially flat, 
although in some cases there is certain evidence for flaring 
or warping (Jura \etal 2007b, 2009). Spectral fitting also finds 
the disks to be predominantly optically thick (Jura 2003; Farihi 
\etal 2010), 
even though this statement may not hold in some parts of the disk 
(Jura \etal 2007b, 2009). These properties make circum-WD debris 
disks look very similar to the ring of Saturn (Cuzzi 2010).

Outer radii $R_{out}$ of these disks are always found to be 
$\lesssim 1$ R$_\odot$ (Jura \etal 2007; Farihi \etal 2010). 
This is close to the Roche radius 
$R_R\sim R_\odot$, within which any object of 
normal density ($\rho\sim 1$ g cm$^{-3}$) would be disrupted by the 
tidal forces of the WD. This observation motivated the idea (Jura 2003)
that compact debris disks owe their origin to the tidal destruction of massive 
asteroid-like bodies, which were scattered into the low-periastron 
orbits by planets that have survived the AGB stage of stellar evolution.

Inner radii of disks are usually small, $R_{in}\sim (15-40)R_\star$, 
where $R_\star$ is the WD radius, although the exact number
depends on the details of spectral modeling (Jura \etal 2007). In 
some cases $R_{in}$ is close to 
the distance at which a refractory body heated by the direct 
starlight would have a temperature equal to the sublimation 
temperature $T_s$, suggesting that the inner cutoff of the disk
is caused by particle sublimation. However, in some cases such 
interpretation implies very high value of $T_s$ (Jura \etal 2007; 
Brinkworth \etal 2009; Dufour \etal 2010). Nevertheless, it is
always safe to assume that the ratio of the outer to inner disk 
radii is rather modest, $R_{out}/R_{in}\lesssim 5$. This will turn 
out being important later on, see \S\ref{subsect:large_narrow}.

A significant reservoir of refractory material in the immediate 
vicinity of the WD can naturally be responsible for contaminating 
the atmospheres of these WDs by metals (Jura 2003). This idea is
strongly supported by the fact that the infrared excesses due to
compact debris disks have been observed only around metal-rich WDs
with rather high metal accretion rates (Farihi \etal 2009, 2010). 
Alternative model of the 
WD metal pollution by interstellar accretion (Dupuis \etal 1993a) 
runs into problem trying to account for the abundance ratios of 
different elements in the WD atmospheres, in particular, greatly 
suppressed hydrogen abundance (Dupuis \etal 1993b). 

The new model of metal accretion from a {\it circumstellar} 
reservoir of high-Z material raises a natural question: how do 
metals get transported to the WD surface from the debris disk,
which has an inner gap with the radius $R_{in}$? It is reasonable to 
assume that at $R_{in}$ disk particles get converted into metal 
gas by the intense heating due to the WD radiation, which then 
viscously accretes onto the WD. However, the issue of whether 
the debris disk can sustain the high metal accretion rates 
$\dot M_Z$ up to several $\times 10^{10}$ g s$^{-1}$ 
inferred in some systems has not been clear 
for a long time.

Recently Rafikov (2011a; hereafter R11) showed that coupling of 
the disk to stellar radiation via the Poynting-Robertson (PR) 
effect (Burns \etal 1979) can result in metal 
accretion rates $\dot M_Z \sim 10^8$ g s$^{-1}$. Later Rafikov 
(2011b) has also demonstrated that under certain circumstances 
even higher values of $\dot M_Z$ can arise from the interaction 
between the debris disk and metal gas that is produced by 
the sublimation at $R_{in}$. 

The model of radiatively driven accretion of R11 did not cover 
the large scale evolution of the debris disk resulting from the 
PR drag, and made certain assumptions (e.g. high optical depths 
of the disk) which were not rigorously verified. The goal of this 
work is to extend the analysis of R11 and to develop a detailed 
global model of the compact debris disk evolution caused by the PR 
drag.

The paper is organized as follows. In \S\ref{sect:theory} we
outline the basic picture of the PR-driven debris accretion 
and derive master equation (\ref{eq:1}) that describes 
global evolution of the disk. We then explore in 
\S\ref{sect:low} both analytically and numerically 
the evolution of a low mass, optically thin disk of debris. In   
\S\ref{sect:high} we study global evolution of massive, 
optically thick debris disks starting with different initial
spatial distributions of debris around the WD (ring-like, 
\S\ref{subsect:large_narrow}, \ref{subsect:small}, or 
disk-like, \S\ref{subsect:disk}) to see the effect on the 
global disk evolution. We discuss our results and their 
observational implications in \S\ref{sect:disc}.


\section{Description of the model.}
\label{sect:theory}


In the following we consider an axisymmetric disk of particles
extending from $R_{in}$ to $R_{out}$ in radius. The 
inner radius may coincide with the sublimation radius $R_s$, 
at which the effective temperature of particles equals the 
sublimation temperature $T_s$:
\ba 
R_s=\frac{R_\star}{2}
\left(\frac{T_\star}{T_s}\right)^{2}\approx 22~R_\star
T_{\star,4}^2 \left(\frac{1500\mbox{K}}{T_s}\right)^2,
\label{eq:R_S} 
\ea
where $R_\star$ is the WD radius, $T_s\approx 1500$ K for silicate
grains, and $T_{\star,4}\equiv T_\star/(10^4$ K) is the normalized
stellar temperature $T_\star$. Taking $R_\star\approx 0.01R_\odot$
one finds $R_s\approx 0.2$ R$_\odot$, in agreement with
observationally inferred inner radii of compact debris disks (Jura
\etal 2007, 2009a). 

When particles reach the sublimation radius they produce metallic gas,
which joins the gaseous disk extending down to the WD surface. Metal 
accretion onto the WD surface proceeds through this disk. 
Although this metal gas also spreads outward from the sublimation
radius (Melis \etal 2010) and under certain circumstances can 
substantially affect 
debris disk evolution (Rafikov 2011b), in this work, following R11,  
we concentrate only on effects associated with the PR drag. Thus,
here we neglect presence of the gas exterior of $R_s$ and its 
interaction with the debris disk. According to Rafikov (2011b)
this is a valid approximation as long as the viscous timescale
in the gaseous disk is shorter than the time on which this disk 
can be replenished by the sublimation of debris. Then the gas does 
not accumulate at $r\approx R_{in}$ and its density is always 
low enough for the gas drag to not affect the debris disk 
appreciably. 

We characterize the disk at each point by its surface density 
$\Sigma(r)$ and optical depth
\ba 
\tau = \frac{3}{4} \frac{\Sigma}{\rho a}, 
\label{eq:tau}
\ea
where $\rho$ is the bulk density of particles and $a$ is their 
characteristic size. The particle size $a$ is not well constrained 
by the existing observations, but its actual value becomes 
important only for low mass disks with small $\tau$, see 
\S \ref{sect:low}.

Following Friedjung (1985) and R11 we represent irradiation of 
the disk by the WD using a single incidence angle $\alpha$ at each
radius $r$ from the disk (the so-called ``lamp-post'' illumination 
model described in R11):
\ba 
\alpha(r) =\frac{4}{3\pi} \frac{R_{\star}}{r}, 
\label{eq:alpha}
\ea
where $R_{\star}$ is the radius of the star. We can then 
introduce {\it optical depth to incoming starlight}
$\tau_\parallel$ according to the following definition:
\ba 
\tau_{\parallel} =\alpha^{-1}\tau. 
\label{eq:tau_par}
\ea
This variable is very important for our subsequent analysis.
In the following we will call the disk optically thick or thin
based on whether $\tpar$ and not $\tau$ is greater or smaller 
than unity.


\subsection{Accretion through the disk.}
\label{subsect:Mdot}

R11 demonstrated that the mass accretion rate $\dot M$ through the
disk of solids driven by the PR drag may be written as
\ba 
\dot M(r)= \alpha(r) \phi_r\frac{L_{\star}}{c^2}. 
\label{eq:mdot}
\ea
Here $L_{\star}$ is the WD luminosity, $c$ is the 
speed of light and function
\ba 
\phi_r = 1 - e^{-\tau_{\parallel}}
\label{eq:phi}
\ea
gives the fraction of incoming starlight that is
absorbed by the disk. 

It is instructive to examine different limits of the expression 
(\ref{eq:mdot}). If the disk is optically thick to incident 
stellar radiation, $\tau_\parallel\gg 1$ (which cannot be the 
case everywhere at all times, as we will see below), 
then the accretion rate 
is independent of $\tau_\parallel$ and is given simply by
\ba 
\dot M_{\tau_{\parallel}\gg 1}(r)\equiv\dot M(r,\tau_\parallel \gg 1)= 
\alpha(r)\frac{L_{\star}}{c^2}.
\label{eq:mdot_PR1}
\ea
In particular, at the sublimation radius given by equation 
(\ref{eq:R_S}) this accretion rate is equal to 
\ba
\dot M_\infty &=& \dot M_{\tau_{\parallel}\gg 1}(R_s)=
\frac{32}{3}  \sigma \left( \frac{R_\star
T_\star T_s}{c}\right)^2 
\label{eq:mdot_PR}\\
&\approx & 7\times 10^7 ~\textrm{g s}^{-1} \left(R_{\star,-2}
T_{\star,4}\frac{T_s}{1500\mbox{K}}\right)^2. 
\ea
R11 has shown this numerical estimate to be consistent
with the lower envelope of the $\dot M_Z$ values inferred for 
the metal-rich WDs exhibiting detectable IR emission 
associated with their debris disks. 

In the opposite limit of a disk, which is optically thin to
incident stellar radiation, $\tau_{\parallel} \ll 1$, one finds
\ba 
\dot M_{\tau_{\parallel}\ll 1} = \tau \frac{L_\star}{c^2}. 
\label{eq:mdot_de}
\ea

It is easy to show that this expression coincides with 
$\dot M$ that one would calculate by simply assuming disk 
particles to be directly illuminated by starlight (i.e. 
unobscured by other particles) and considering each of them 
as independent from others. Azimuthal PR drag force acting on 
a single, perfectly absorbing particle of mass 
$m=(4\pi/3)\rho a^3$ is given by
\ba 
F_\varphi = \frac{L_\star}{4\pi r^2} \pi a^2 \frac{\Omega_K r}{c^2}, 
\ea
where $\Omega_K$ is Keplerian angular velocity. This force gives 
rise to radial drift velocity
\ba 
v_r = \frac{2F_\varphi}{m\Omega_K} = 
\frac{3}{8\pi}\frac{L_\star}{\rho a c^2 r}, 
\ea
and results in mass accretion rate $\dot M=2\pi r v_r\Sigma$, which 
is easily shown to reduce to equation (\ref{eq:mdot_de}).


\subsection{Evolution equations}
\label{subsect:evol}

Evolution of the debris disk is described by the continuity equation 
\ba 
\frac{\partial \Sigma}{\partial t} - \frac{1}{2\pi r}
\frac{\partial \dot M}{\partial r} = 0,
\label{eq:cont}
\ea
with $\dot M$ given by equation (\ref{eq:mdot}).

We introduce new dimensionless time and space variables
\ba 
x \equiv \frac{r}{R_{in}}, \qquad T \equiv \frac{t}{t_0}, 
\label{eq:dim}
\ea
where 
\ba 
t_0 \equiv \frac{8\pi}{3} \frac{\rho a R_{in}^2 c^2}{L_{\star}}
\label{eq:t_0}
\ea
is the characteristic timescale of the problem. If $R_{in}$ coincides
with the sublimation radius $R_s$ defined in (\ref{eq:R_S}) then
\ba 
t_0 = \frac{1}{6} \frac{\rho a c^2}{\sigma T_s^4}
\approx 5 \times 10^4 \textrm{ yr }~\frac{a}{1 \textrm{
cm}}\left(\frac{1500~\mbox{K}}{T_s}\right)^{4},
\label{eq:t_0_sub}
\ea
where we took $\rho = 3$ g cm$^{-3}$.

Using equations (\ref{eq:alpha})-(\ref{eq:phi}) and definitions
(\ref{eq:dim}), (\ref{eq:t_0}) we can bring the continuity 
equation (\ref{eq:cont}) to a scale-free form:
\ba 
\frac{\partial \tau_{\parallel}}{\partial T} -
\frac{\partial}{\partial x} \left(
\frac{1-e^{-\tau_{\parallel}}}{x} \right) = 0. 
\label{eq:1}
\ea
A notable fact about equation (\ref{eq:1}) is that it does 
not contain any free parameters. 

Once $\tpar(x,T)$ is obtained by solving this equation, one can 
easily infer mass accretion rate at any point in the disk since
\ba 
\dot M(x,T) 
=\dot M_{\infty}\frac{1-e^{-\tau_\parallel}}{x},
\label{eq:y}
\ea
where $\dot M_{\infty}$ is defined by equation (\ref{eq:mdot_PR}). 
Because the viscous timescale of the gas produced
by debris sublimation is very short compared to the disk evolution
time (R11), the metal accretion rate onto the WD surface 
$\dot M_Z$ is given simply by
\ba
\dot M_Z(T)=\dot M(x=1,T)=\dot M_{\infty}
\left[1-e^{-\tau_\parallel(x=1,T)}\right].
\ea
In the following we will use 
$\dot M_Z(T)$ and $\dot M(x=1,T)$
interchangingly. 

Analysis of equation (\ref{eq:1}) may be simplified if 
we introduce a new function $y(x,T)\equiv 
\dot M(x,T)/\dot M_{\infty}=\left(1-e^{-\tau_\parallel}\right)/x$,
which is just the mass accretion rate through the debris disk 
normalized by $\dot M_{\infty}$. Then equation (\ref{eq:1}) transforms to
\ba 
\frac{\partial y}{\partial T} + \left( y - \frac{1}{x} \right)
\frac{\partial y}{\partial x} = 0. 
\label{eq:2}
\ea
It has an implicit solution 
\ba 
y = f(xy - y^2 T + \ln(1-xy)), 
\label{eq:imp} 
\ea
where $f$ is an arbitrary function, which is set by the initial 
surface density distribution of the debris. Unfortunately, 
inferring $f$ from $y(x,T=0)$ is not a trivial task, which makes
analysis of solution (\ref{eq:imp}) very difficult. Nevertheless, 
we can still deduce useful information about the behavior of
this solution in some specific limits, see \S \ref{sect:low}
and Appendix \ref{app:1}. 

We now proceed to investigate the details of the global disk 
evolution with different initial conditions.


\section{Low mass disks.}
\label{sect:low}


We start by considering the case of a tenuous disk, in which the 
optical depth for incident stellar radiation $\tau_\parallel\ll 1$.
In that case according to equation (\ref{eq:y}) 
$y\approx \tau_\parallel/x\ll x^{-1}$ and the nonlinear 
term $ydy/dx$ can be neglected in equation (\ref{eq:2}) reducing it
to 
\ba 
\frac{\partial y}{\partial T}  - \frac{1}{x} 
\frac{\partial y}{\partial x} = 0. 
\label{eq:lin}
\ea
This linear equation has an explicit solution 
\ba 
\frac{\tau_\parallel}{x}\approx y = y_0(\sqrt{x^2+2T}),
\label{eq:sol}
\ea
where $y_0(x)=y(x,T=0)\approx \tau_\parallel(x,T=0)/x$ 
is the initial spatial distribution 
of variable $y$ set by the initial 
surface density distribution $\tau_\parallel(x,T=0)$. 

If initially the debris was concentrated in a structure (a disk 
or a ring) with characteristic dimensionless scale $x$ (physical 
scale $R_{in}x$), then it follows from equation (\ref{eq:lin}) 
that the characteristic time $t_{thin}$ on which $\tau_\parallel$ 
evolves is
\ba 
t_{thin}=t_0 x^2,~~~~~\mbox{or}~~~~~T_{thin}=x^2.
\label{eq:T_thin}
\ea
This time is independent of the disk mass but is sensitive to 
its size $x$ and particle properties, see equation (\ref{eq:t_0_sub}). 
This is not surprising since when $\tau_\parallel\ll 1$ particles 
interact with the WD radiation independently of each other 
and drift towards the WD on a PR timescale of an {\it individual} 
particle. The latter coincides with $t_{thin}$ up to constant 
factors and is indeed a function of $\rho$, $a$ and the distance $r$ 
that needs to be traveled.

\begin{figure}[h]
\begin{center}
\includegraphics[width=0.95\linewidth]{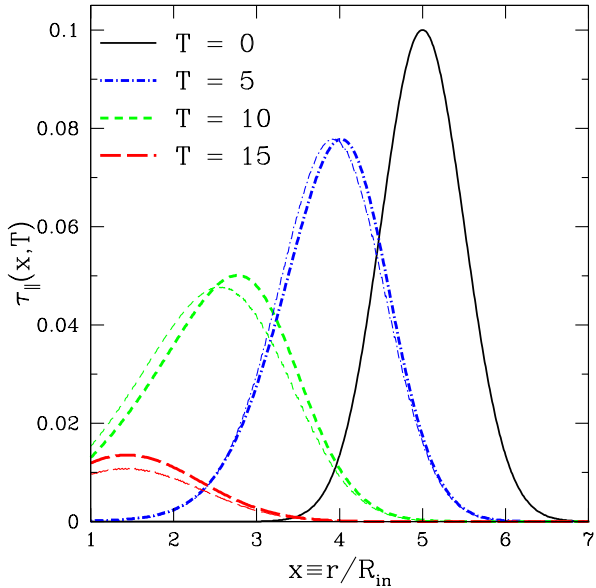}
\caption{Evolution of a low mass debris disk around a WD with initial 
$\tau_\parallel\ll 1$ everywhere. Initial density profile is given 
by equation (\ref{eq:prof}) with $\tau_{\parallel,0} = 0.1$, 
$x_0 = 5$, $\sigma = 0.7$. Thick curves are the numerical results 
for the optical depth to incoming starlight $\tpar(x,T)$
(related to $\Sigma(r,t)$ via equations 
(\ref{eq:tau})-(\ref{eq:tau_par}))
at different moments of time indicated on the panel, while the thin 
curves represent analytical solution (\ref{eq:thin_sol}).
\label{fig:thin}} 
\end{center}
\end{figure}

Disks which are optically thin to incident stellar radiation must 
satisfy a certain mass constraint. The mass of a disk extending out
to $R_{out}=x_{out}R_{in}$ can be written as
\ba
M_d &=& 2\pi\int\limits_{R_{in}}^{R_{out}}\Sigma(r)rdr=
M_0\int\limits_1^{x_{out}}\tpar(x)dx,
\label{eq:M_d}
\ea
where the characteristic disk mass $M_0$ is defined as
\ba
M_0\equiv \frac{32}{9}\rho a R_{\star} R_{in}\approx
10^{20}~ \textrm{g}~\frac{a}{1 \textrm{ cm}}~\frac{R_{in}}
{0.2 R_{\odot}}.
\label{eq:M_0}
\ea
and the numerical estimate assumes $\rho = 3$ g cm$^{-3}$, and 
$R_\star \approx 0.01 R_{\odot}$. This numerical value of $M_0$
corresponds to the disruption of an asteroid with the diameter of 
40 km. 

The disk which is optically 
thin everywhere ($\tpar\lesssim 1$) should have mass 
$M_d\lesssim M_0$ since $x_{out}-1$ is of order unity for WD 
debris disks.


\subsection{Numerical results.}
\label{subsect:thin-num}

We numerically integrated equation (\ref{eq:1}) for an initial
ring-like density profile given by 
\ba 
\tau_{\parallel}(x,0)\approx xy_0(x) = \tau_{\parallel,0}
\exp\left[-\frac{(x-x_0)^2}{\sigma^2}\right], 
\label{eq:prof}
\ea
with $\tau_{\parallel,0} \ll 1$.
In Figure \ref{fig:thin} we show the results obtained for the case
$\tau_{\parallel,0} = 0.1$, $x_0 = 5$, $\sigma = 0.7$, i.e. a rather 
narrow, tenuous ring, with the radius large compared to $R_{in}$.
We also plot analytical solution for $\tau_{\parallel,0} \ll 1$
\ba 
\tau_{\parallel}(x,T) \approx \frac{\tau_{\parallel,0}x}{\sqrt{x^2+2T}}
\exp\left[-\frac{(\sqrt{x^2+2T}-x_0)^2}{\sigma^2}\right], 
\label{eq:thin_sol}
\ea
which follows from (\ref{eq:sol}). One can see that analytical 
solution reproduces the numerical results quite well. As time goes 
by the ring of material drifts towards the WD and broadens, simply
because the speed of inward migration due to the PR drag increases 
as $x$ decreases. This eventually results in substantial surface 
density reaching the inner radius and giving rise to sublimation there. 
Metal gas produced at $R_{in}$ then directly accretes onto the WD.

\begin{figure}[h]
\begin{center}
\includegraphics[width=0.95\linewidth]{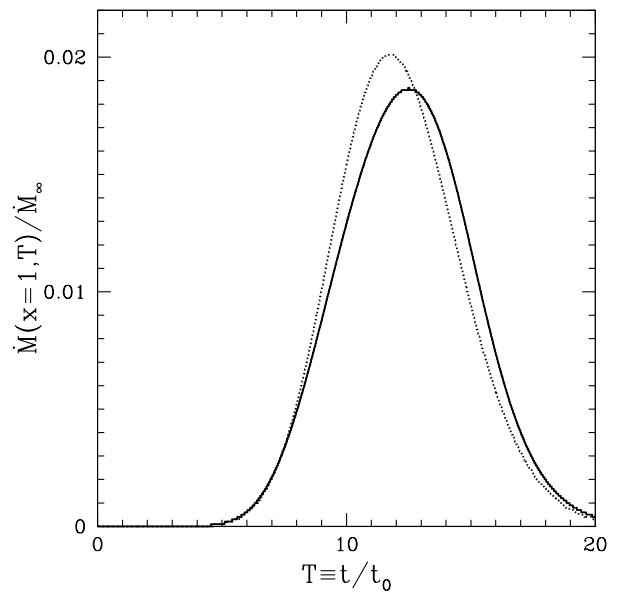}
\caption{Metal accretion rate at $r=R_{in}$ for a low-mass optically 
thin disk (same as in Figure \ref{fig:thin}), normalized 
by $\dot M_{\infty}$. Solid and dotted curves are the numerical and 
analytical [equation (\ref{eq:mdot_thin})] results.
\label{fig:mdot1}}
 \end{center}
\end{figure}

Metal accretion rate $\dot M_Z\ll \dot M_\infty$ in the optically 
thin case, simply because the initial value of $\tpar\ll 1$ 
everywhere in the disk and equation 
(\ref{eq:sol}) then predicts that $y(x=1,T)\ll 1$ for any $T$.
Using solution (\ref{eq:sol}) we can derive an explicit formula for 
$\dot M(x=1,T)$ for the initial debris density distribution 
(\ref{eq:prof}):
\ba 
\dot M(x=1,T) \simeq 
\frac{\dot M_\infty \tau_{\parallel,0}}{\sqrt{1+2T}} 
 \exp\left[-\frac{(\sqrt{1+2T}-x_0)^2}{\sigma^2}\right]. 
\label{eq:mdot_thin}
\ea 
In Figure \ref{fig:mdot1} we compare this expression with the numerically
derived $\dot M(x=1,T)$ and again the agreement is reasonably good. 
Mass accretion rate peaks at $T\approx x_0^2/2\approx 12$ when the 
expression inside the exponential in (\ref{eq:mdot_thin}) goes to zero.
The maximum accretion rate achieved at this moment is about  
$\dot M_\infty\tau_{\parallel,0}/x_0\approx 0.02\dot M_\infty$,
in agreement the numerical results shown in Figure \ref{fig:mdot1}.

\section{High mass disks.}
\label{sect:high}


Now we consider evolution of massive debris disks with mass exceeding 
$M_0$ defined by (\ref{eq:M_0}). Such disks have $\tau_\parallel\gg 1$ 
at least at some radii. This condition 
cannot be true throughout the whole disk as there should always be  
regions in which $\tau_\parallel\lesssim 1$, e.g. near the disk 
edges. Even if they do not exist initially such optically thin 
regions naturally appear as a result of disk evolution as we 
show below. 

Whenever $\tau_\parallel\gg 1$ one can neglect the term proportional 
to $e^{-\tau_\parallel}$ in equation (\ref{eq:1}) and obtain the
following simple solution:
\ba
\tau_\parallel(x,T)=\tau_\parallel(x,T=0)-\frac{T}{x^2},
\label{eq:thick_sol}
\ea
or, in dimensional units,
\ba
\Sigma(r,t)=\Sigma(r,t=0)-\frac{2}{3\pi^2}\frac{R_\star}{r^3}
\frac{L_\star}{c^2}t.
\label{eq:thick_sol_sigma}
\ea
This solution implies that in parts of the disk, which are optically 
thick to incoming stellar radiation the surface density steadily 
decreases in time at a constant rate set only by the distance from 
the WD. This behavior is a direct consequence of the $\dot M$ saturation
at the level $\dot M_{\tpar\gg 1}$ independent of $\tau_\parallel$ 
in the case $\tau_\parallel\gg 1$, see equation (\ref{eq:mdot_PR1}).

At a glance this kind of evolution looks very strange as the 
debris does not seem to move anywhere --- $\Sigma$ simply goes 
down at all radii. This interesting behavior can be understood 
only by accounting for the 
optically thin regions that should inevitably be present 
in the disk and their interplay with the optically thick part 
of the disk. We demonstrate this in \S \ref{subsect:thick-num}.

Solution (\ref{eq:thick_sol}) also implies that the characteristic 
timescale of the disk evolution is now
\ba  
t_{thick} \sim  t_0\tau_{\parallel}(x)x^2=\frac{2\pi\Sigma(r) r^2}
{\dot M_{\tpar\gg 1}(r)}, 
\label{eq:T_thick}
\ea
or $T_{thick}=\tau_{\parallel}(x)x^2$ in dimensionless units.
Clearly, $t_{thick}$ is (up to factors of order unity) just the time, 
on which the characteristic disk mass $\sim \Sigma(r)r^2$ gets 
exhausted by accretion at the rate $\dot M_{\tpar\gg 1}(r)$. 
Unlike $t_{thin}$ defined by equation (\ref{eq:T_thin}) $t_{thick}$ 
is independent of particle properties --- $\rho$ and $a$, but
scales linearly with the disk mass.


\subsection{Numerical results.}
\label{subsect:thick-num}

We numerically integrated evolution of several initially 
optically thick density distributions. The morphologies we
consider range from a narrow ring to an extended disk of debris.
They represent different initial spatial distributions
of the debris that might possibly result from the tidal disruption
of an asteroid. We choose all disks to have the same total mass
$M_d=350M_0$ initially and concentrate on exploring the differences of 
their evolutionary routes caused purely by the morphology.


\subsubsection{Large and narrow ring.}
\label{subsect:large_narrow}

\begin{figure}[h]
  \begin{center}
 \includegraphics[width=0.95\linewidth]{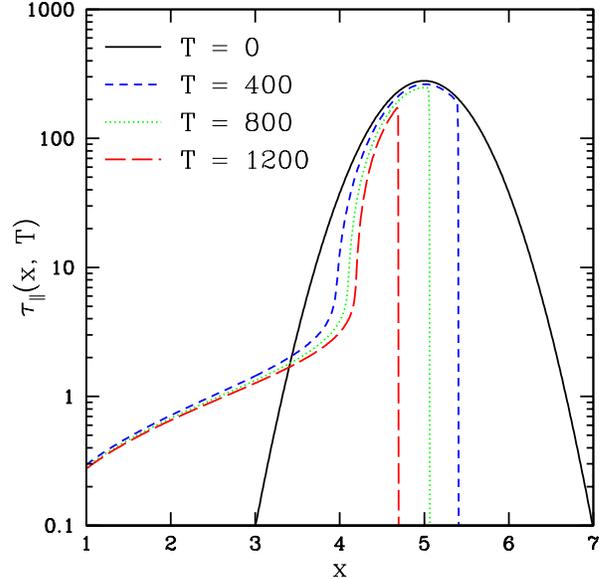}
 \caption{Evolution of $\tpar$ for an optically thick 
disk with the gaussian initial density profile (\ref{eq:prof}), 
assuming $x_0 = 5$, 
$\sigma = 0.7$ and total mass $M_d=350M_0$ (large and narrow ring). Radial 
profiles of $\tpar$ are shown at different moments of time 
indicated on the panel. Note the appearance of an optically thin 
tail of debris extending to $x=1$ and the sharpness of the outer disk 
edge.
\label{fig:gauss5}}
 \end{center}
\end{figure}

We start by exploring the initial distribution in the form (\ref{eq:prof})
with $x_0=5$, $\sigma=0.7$ but now we take $\tau_{\parallel,0}=280$ so 
that the disk mass $M_d=350M_0\gg M_0$. Resulting evolution of
$\tau_\parallel$ is shown in Figure \ref{fig:gauss5}. Apart from the 
theoretically expected steady decrease of $\tau_\parallel$ with time 
in the bulk of the disk where $\tau_\parallel\gg 1$ one immediately 
notices two interesting features of the solutions: (1) a sharp drop of 
the disk surface density at the outer edge of the disk, and (2) an 
optically thin 
($\tau_\parallel\lesssim 1$) tail of debris extending from the optically 
thick region all the way to $R_{in}$, which remains stable throughout
the evolution of the disk. We now discuss the nature of these features 
in more detail.

Sharp outer edge appears because the evolution timescale of a particular 
disk region $t_{thick}\propto\tpar$, see equation (\ref{eq:T_thick}).
Initial density distribution in the form (\ref{eq:prof}) has $\tpar$
increasing {\it inwards} in the outer part of the ring. This causes 
the outermost extremity of the ring (which is optically thin
and evolves on timescale $t_{thin}$) to migrate towards the WD faster 
than the parts of the ring closer to the WD do. As a result, the 
outermost disk material catches up with the debris at smaller $x$ having 
higher $\tpar$, leading to a debris pileup there and formation of a 
sharp outer edge. This edge appears on a timescale 
$\sim t_{thin}$ since this is the characteristic time, on which 
the material from the optically thin outer disk region reaches 
$x\approx x_0$ where most of the ring mass is concentrated.
We provide a more rigorous justification for the outer edge 
appearance in Appendix \ref{app:1}.

\begin{figure}[h]
  \begin{center}
 \includegraphics[width=0.95\linewidth]{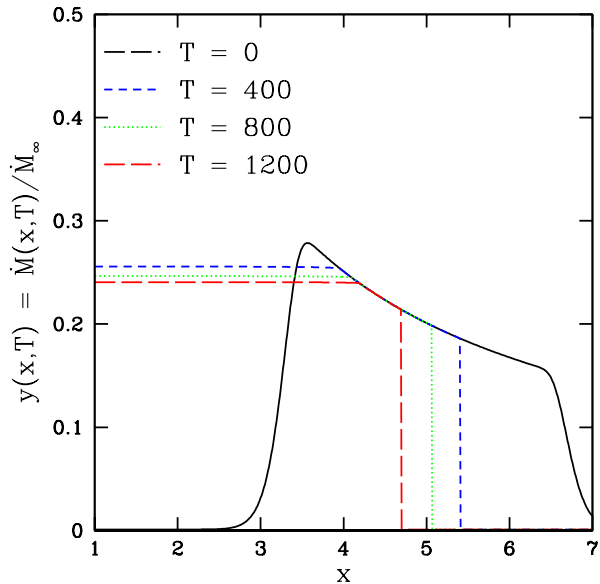}
 \caption{Radial distribution of the dimensionless mass accretion rate
 $y(x,T)=\dot M(x,T)/\dot M_\infty$ (equation (\ref{eq:y})) 
through the debris disk at different moments of time. Results are
shown for the same disk as in Fig. \ref{fig:gauss5}. Note that 
$y \to const$ near $x=1$, i.e. in the optically thin tail.
\label{fig:y}}
 \end{center}
\end{figure}

Origin of the quasi-stationary tail with $\tau_\parallel\lesssim 1$ 
at small $r$ can be understood as follows. Even if initially the disk 
has $\tpar\gg 1$ everywhere such tail would rapidly form since, 
according to equation (\ref{eq:thick_sol}), at any distance from the 
WD $\tpar$ should switch into the optically thin regime 
within finite time. 
This happens fastest at the inner edge of the disk. 
After the tail has formed it evolves on 
a timescale $t_{thin}$ given by equation (\ref{eq:T_thin}). 
Inner parts of the tail evolve more rapidly than the 
outer ones and a quasi-steady state is attained in the tail with 
$\dot M$ almost independent of $r$. This point is illustrated in 
Figure \ref{fig:y}, where we show the run of $\dot M(x)/\dot M_\infty$
for a calculation displayed in Figure \ref{fig:gauss5}
at different moments of time. One can easily see that after some 
evolution has taken place $\dot M(x)$ indeed 
becomes independent of $x$ 
in the tail. 

Thus, the behavior of 
$\tpar(x,T)$ in the tail can be described by the equation
(cf. equation (\ref{eq:y}))
\ba 
\frac{1-\exp\left[-\tau_\parallel(x,T)\right]}{x}= y_{tail}(T)=
\frac{\dot M_{tail}(T)}{\dot M_\infty}, 
\label{eq:tail}
\ea
where $\dot M_{tail}(t)$ is the mass accretion rate through the tail
(and $y_{tail}$ is the dimensionless analog of this quantity), which 
in general is a function of time. Equation (\ref{eq:tail}) describes 
the behavior of $\tpar$ in Figure 
\ref{fig:gauss5} as a function of $x$ quite well. 

At some radius $x_{\tau=1}=r_{\tau=1}/R_{in}$ the tail becomes 
optically thick, i.e. $\tpar(x_{\tau=1})=1$. A this 
point the tail merges with the optically thick part of the disk and
the evolution timescale, which is now given by $t_{thick}$, 
rapidly increases, see equation (\ref{eq:T_thick}). Accretion rate 
through the tail $\dot M_{tail}(T)$ is set almost exclusively 
by the value of $x_{\tau=1}$ since 
by definition $1-\exp\left[-\tau_\parallel(x_{\tau=1},T)\right]\sim 1$ 
independent of $T$, so that
\ba
\frac{\dot M_{tail}(T)}{\dot M_\infty}\approx \frac{1}{x_{\tau=1}(T)}
\label{eq:dotM_tail}
\ea
at all times. Thus, it is the dependence of $x_{\tau=1}$ on $T$ that
determines how the mass accretion rate through the tail and at 
$R_{in}$ changes with time.

To find $x_{\tau=1}(T)$ we need to set the left hand side of equation
(\ref{eq:thick_sol}) to unity and solve the resultant equation for
$x$ as a function of $T$. Since here we consider the case of 
$\tpar(x,T=0)\gg 1$ the relation between $x_{\tau=1}$ and $T$ is provided  
with sufficient accuracy by solving the equation
\ba
\tau_\parallel(x_{\tau=1},T=0)=\frac{T}{x_{\tau=1}^2},
\label{eq:thick_sol1}
\ea
for a given initial density distribution in the disk.

\begin{figure}[h]
  \begin{center}
 \includegraphics[width=0.95\linewidth]{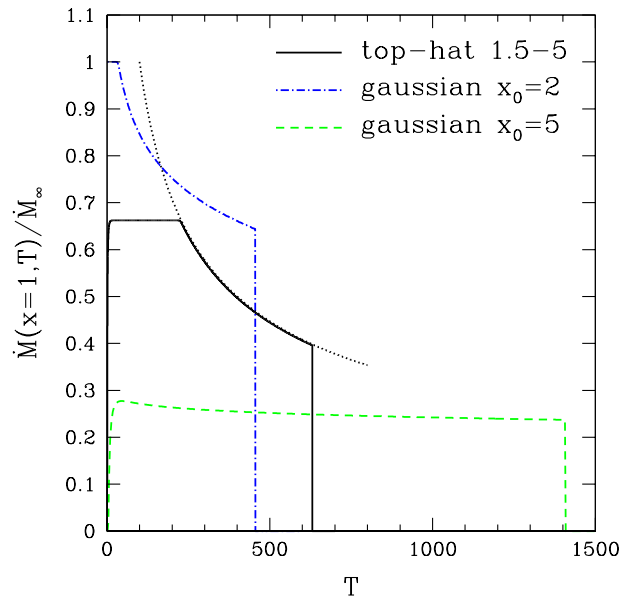}
 \caption{Mass accretion rate at the inner disk radius $\dot M(x=1,T)$ 
normalized by $\dot M_\infty$ for different initial distributions of 
the debris surface density: large narrow ring (\S 
\ref{subsect:large_narrow}; dashed line), small ring  (\S 
\ref{subsect:small}; dot-dashed line), and extended top-hat disk  (\S 
\ref{subsect:disk}; solid line). Dotted curve shows analytical
prediction (\ref{eq:approx}) for $\dot M_Z$ in the top-hat case. 
Disk mass is the same in all three cases: $M_d=350M_0$.
\label{fig:mdot2}}
 \end{center}
\end{figure}

For the Gaussian initial distribution in the form (\ref{eq:prof}) 
one finds that
\ba
x_{\tau=1}=x_0-\sigma
\left(\ln\frac{\tau_{\parallel,0}x_{\tau=1}^2}{T}\right)^{1/2}
\label{eq:x_cr}
\ea
Whenever $x_0\gg 1$ and $\sigma\ll x_0$, which is a reasonable 
approximation for the case shown in Figure \ref{fig:gauss5},  
equation (\ref{eq:x_cr}) predicts 
that $x_{\tau=1}\approx x_0$ and that $x_{\tau=1}$ depends on 
time $T$ only weakly (logarithmically). This explains 
the time invariance of the optically thin tail in Figure 
\ref{fig:gauss5} and only weak time dependence of $\dot M$ in the 
tail in Figure \ref{fig:y}: the mass flux through the 
tail 
\ba
\dot M_{tail}\approx \dot M_\infty x_0^{-1} 
\label{eq:dotMtail}
\ea
is almost independent of $T$ if the 
initial mass distribution has the form of a large and narrow 
ring (i.e. $\sigma\ll x_0$). As a result, the flux of metals 
onto the WD $\dot M(x=1,T)$ also varies with time only weakly
for ring-like initial distribution of the debris, as can be 
seen in Figure \ref{fig:mdot2}. For debris 
disks around WDs produced by tidal disruption of asteroids 
$x_0\lesssim 5$ and so equation (\ref{eq:dotMtail}) 
guarantees that $\dot M_Z$ stays above 
$0.2\dot M_\infty$ or so, see Figure \ref{fig:mdot2}. 

Equation (\ref{eq:tail}) also predicts that $\tpar$ does not 
go significantly below unity even at the inner edge of the disk:
$\tpar(x=1,T)\approx x_0^{-1}$ depends only on $x_0$, see 
equation (\ref{eq:tail}), and does not change appreciably with 
time for a narrow ring.  

It is clear that the general evolutionary picture presented 
in Figure \ref{fig:gauss5}
would still hold if we assume a ring density profile other than 
Gaussian, as long as the width of the ring is small and its radius 
is large compared to $R_{in}$.


\subsubsection{Small ring.}
\label{subsect:small}
    
\begin{figure}[h]
\begin{center}
 \includegraphics[width=0.95\linewidth]{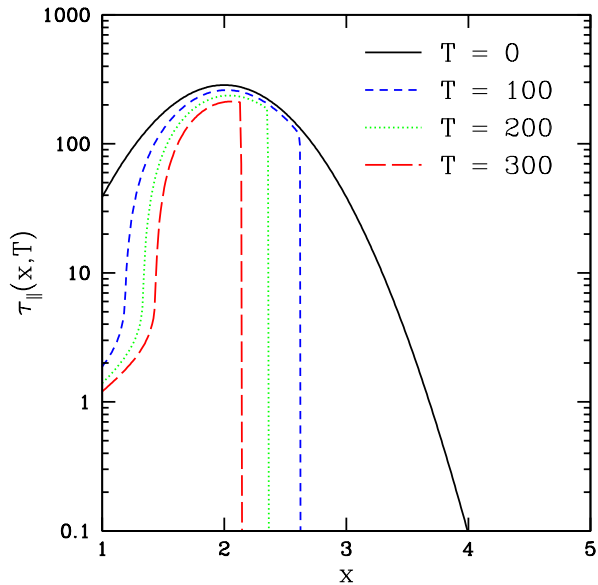}
 \caption{Same as Figure \ref{fig:gauss5} but for the gaussian 
initial profile (\ref{eq:prof}) 
with $x_0 = 2$, $\sigma = 0.7$ and total disk mass $M_d = 350M_0$ 
(small ring). 
\label{fig:gauss2}}
 \end{center}
\end{figure}

We now consider evolution of initial distribution in the Gaussian 
form (\ref{eq:prof}) but with $x_0 = 2$, $\sigma = 0.7$, and truncated
at $x=1$ (i.e. $\tpar=0$ for $x<1$), while keeping the total disk 
mass the same, $M_d = 350M_0$. In this case the width of the ring is 
no longer small compared to its radius. Snapshots of the $\tpar$
profile at different time $T$ are presented in Figure \ref{fig:gauss2}.

Since $x_0-1\sim \sigma$ this distribution has significant 
$\tpar\approx 40$ at $x=1$ at the very beginning. According 
to equation (\ref{eq:thick_sol1}) it takes time $T\approx 
\tau_{\parallel,0}(x=1)\approx 40$ for $\tpar(x=1)$ to drop
to small value $\sim 1$, and during this initial stage 
$\dot M(x=1,T)\approx \dot M_\infty$, as Figure \ref{fig:mdot2}
clearly shows. 

Beyond this point a marginally optically thin tail develops near
$x=1$ but it never reaches a steady state. This can be understood by 
inspection of the transcendental equation (\ref{eq:x_cr}) determining 
the crossover radius $x_{\tau=1}$ (and the $\dot M$ through the 
tail): whenever the width of the ring $\sigma$ is not small
compared to its radius $x_0$ this equation does not have 
a time-independent solution. As a result, $\dot M_Z=\dot M(x=1,T)$ 
noticeably varies in time, see Figure \ref{fig:mdot2}.

Also note that the sharp outer edge of the disk is again present
in Figure \ref{fig:gauss2} --- its origin is the same as in the 
case of the large and narrow ring explored in \S 
\ref{subsect:large_narrow}.

\subsubsection{Extended disk.}
\label{subsect:disk}

\begin{figure}[h]
\begin{center}
\includegraphics[width=0.95\linewidth]{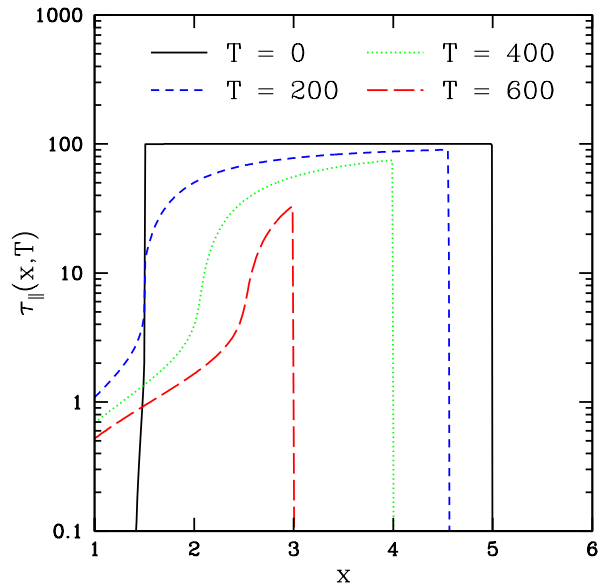}
\caption{Same as Figure \ref{fig:gauss5} but for a ``top-hat''
initial density distribution $\tau_{\parallel}(x,T=0)=100$ for $1.5<x<5$.
Note that both the optically thin tail near $x=1$ and the sharp 
outer edge are still present.
\label{fig:tophat}}
\end{center}
\end{figure}

Finally we explore the case of an extended initial density distribution 
in the form of a disk described by a ``top-hat'' initial condition
\ba
\tau_\parallel(x,T=0) = \tau_{\parallel,0}=100,~~~ x_1<x<x_2, 
\label{eq:top-hat}
\ea
where $x_1=1.5$ and $x_2=5$. Evolution of this distribution is 
shown in Figure \ref{fig:tophat} and one can again clearly see 
the development of an optically thin tail at the inner edge of 
the disk and the sharp outer edge. 

The tail appears on a timescale $T_{thin}(x_1)\approx x_1^2\approx 2$.
Initially this tail attaches directly to $x_{\tau=1}=x_1$ because 
it takes time $T_{thick}(x_1)\approx \tau_{\parallel,0}x_1^2
\approx 220$ (see equation (\ref{eq:thick_sol1})) for $\tpar(x_1)$ 
to go down to unity. During this period of time $\dot M_Z\approx 
x_1^{-1}\dot M_\infty\approx 0.66 M_\infty$ stays virtually constant.
These simple analytical conclusions are in excellent agreement 
with the numerical results presented in Figure \ref{fig:mdot2}.

After $\tpar(x_1)$ has dropped below unity the crossover radius 
$x_{\tau=1}$ starts moving out and $\dot M_{tail}$ goes down. Using
equations (\ref{eq:thick_sol1}) and (\ref{eq:top-hat}) we infer that 
in the top-hat case $x_{\tau=1}\approx (T/\tau_{\parallel,0})^{1/2}$ 
for $T>T_{thick}(x_1)$, resulting in
\ba
\dot M_Z\approx 
\dot M_\infty \left(\frac{\tau_{\parallel,0}}{T}\right)^{1/2}. 
\label{eq:approx}
\ea
This analytical prediction is shown by dotted line in 
Figure \ref{fig:mdot2} and clearly agrees well with the 
numerical result for the initial top-hat density distribution.

\section{Discussion.}
\label{sect:disc}

In this paper we focused on studying how the debris disk evolves 
due to the PR drag alone. Among other radiative effects potentially 
affecting such disks we can mention the direct transfer of the radiative 
angular momentum to the disk that occurs if the WD is rapidly rotating. 
Previously Walker \& Meszaros (1989) suggested this process to be 
important for the evolution of accretion disks around neutron stars 
experiencing Type I X-ray bursts, which often rotate quite rapidly. 
In the case of WDs there is no good evidence for rapid rotation
(Kawaler 2004) with several disk-hosting systems showing 
$v_{rot}\sin i\lesssim 20$ km s$^{-1}$ (G\"ansicke \etal 2006, 2007). 
However, 
one can easily demonstrate that even if WD were to rotate at 
breakup speed the radiative angular momentum transport would still 
not produce WD disk evolution comparable to that caused by the 
PR drag. 

Indeed, the angular momentum carried by photons emitted by the WD
rotating with equatorial speed $v_{rot}$ can be estimated as 
$H\sim (L_\star/c^2)R_\star v_{rot}$. Disk absorbs some fraction
of this angular momentum which exerts radiative torque 
${\cal T}_{rad}$ on the disk. The ratio of this torque 
to the torque due to the PR drag ${\cal T}_{PR}$ is easily 
shown to be
\ba
\frac{{\cal T}_{rad}}{{\cal T}_{PR}}\sim \frac{R_\star v_{rot}}
{\Omega_K r^2},
\label{eq:Trat}
\ea 
where $\Omega_K$ is the Keplerian angular frequency. Even if the WD 
rotates at breakup and $v_{rot}\sim \Omega_K(R_\star)R_\star$ one still 
finds ${\cal T}_{rad}/{\cal T}_{PR}\sim (R_\star/r)^{1/2}$, which
is less than unity because the debris disk lies quite far from the 
WD surface, i.e. $r\gg R_\star$. Moreover, measured rotational speeds 
of the aforementioned WDs are $\sim 10^{-2}$ of their breakup speed,
so the direct radiative transfer of angular momentum to debris 
has negligible effect on disks in these systems.

One of the key results of this work is that the low metal 
accretion rates onto the WD $\dot M_Z\ll\dot M_\infty$ 
are possible only if the debris disk hosted
by the WD has small mass, below $M_0$ given by equation (\ref{eq:M_0}), 
and is everywhere optically thin to incident stellar radiation.
This is demonstrated in \S \ref{sect:low} where we find 
$\dot M_Z\approx \dot M_\infty\tpar(x=1)\ll  \dot M_\infty$ 
(see equation (\ref{eq:sol})) throughout 
the whole evolution of the low mass disk.

On the contrary, if the disk is massive enough to contain 
optically thick regions, which typically requires 
$M_d\gtrsim 10^{20}-10^{21}$ g, then $\dot M_Z$ is guaranteed to not 
fall below $(R_{in}/R_{out})\dot M_\infty\approx 0.2\dot M_\infty$, as 
we have shown in \S \ref{sect:high}. This is in agreement with the 
statement in R11 that $\dot M_Z$ should not deviate significantly 
from $\dot M_\infty$ for optically thick disks.

\begin{figure}[h]
  \begin{center}
 \includegraphics[width=0.95\linewidth]{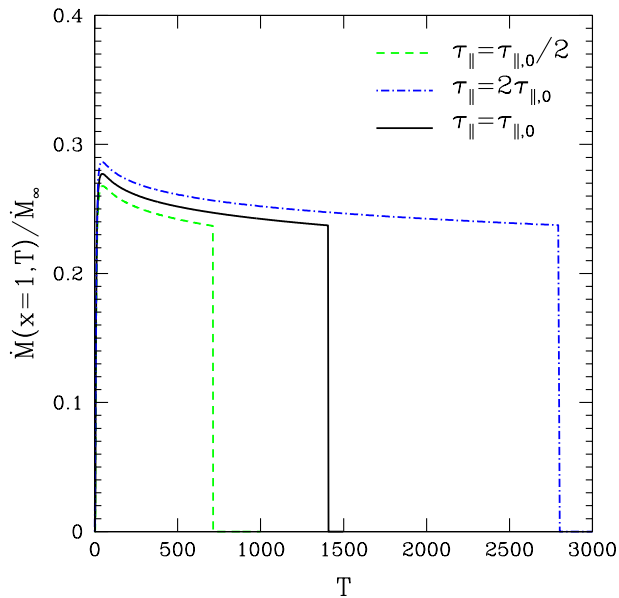}
\caption{Time evolution of the mass accretion rate at the inner 
radius $\dot M(x=1,T)$ for three massive, optically thick disks with 
the same initial gaussian density profile (\ref{eq:prof}) and $x_0 = 5$, 
$\sigma = 0.7$,  
but different initial masses in
the ratio 0.5:1:2. Note that the amplitude of  $\dot M(x=1,T)$ is 
essentially independent of the disk mass, only the disk lifetime 
depends on it.
\label{fig:mdot3}}
 \end{center}
\end{figure}

This lower bound on $\dot M_Z$ for massive disks is very robust, 
in particular it 
does not depend on the disk mass. This point is illustrated in Figure 
\ref{fig:mdot3} where we show the time evolution of $\dot M_Z$ for 
three disks with the same initial Gaussian density distribution 
(\ref{eq:prof}) but with different initial masses $M_d$ set to be in 
the ratio $0.5:1:2$. One can easily see that the only thing, which 
is different between these three cases
is the disk lifetime (which is always proportional to the disk mass),
while $\dot M_Z$ is essentially the same.

The time evolution of $\dot M_Z$ in the optically thick case is 
found to depend predominantly on the shape of the initial density 
distribution. Debris distributions in the form of a narrow ring tend to
produce  $\dot M_Z$ only weakly varying with time. Extended initial 
distributions of the debris with radial width comparable to mean radius 
generally lead to $\dot M_Z$ varying in time and decaying by 
a factor of several by the end of the disk lifetime, see Figure 
\ref{fig:mdot2}.

Based on these results one should expect WDs with
strong IR excesses, signifying massive, optically thick debris disk
around them, to exhibit $\dot M_Z\gtrsim 0.2\dot M_\infty$. This is exactly 
what the current data seem to suggest: no WD with a debris 
disk detected via its IR signature is inferred to have $\dot M_Z$ below 
$\dot M_\infty$ (R11). 

However, if the systems exhibiting IR excesses were 
found to show $\dot M_Z\ll \dot M_\infty$, the debris disks around 
them should be optically thin. It is then
natural to expect a positive correlation between the strength 
of the IR excess and the $\dot M_Z$ in such systems, since both are 
linearly proportional to $\tpar\ll 1$ in the optically thin regime. 
Indeed, $\dot M_Z\propto\tpar$ according to (\ref{eq:sol}) while 
the amount of the WD luminosity that the optically thin disk 
intercepts and reradiates in the near-IR is
\ba
L_d\approx \frac{L_\star}{2}\int\limits_{R_{in}}^{R_{out}}
\frac{\alpha(r)}{r}\tpar(r)dr,
\label{eq:L_d}
\ea
which also scales positively with $\tpar$.

The timescale on which debris disks get depleted can be quite 
different in the 
optically thick and thin cases. Characteristic lifetime 
of an optically thin disk composed of $\sim 1$ cm particles 
$t_{thin}$ is several 
$10^5$ yr, as equations (\ref{eq:t_0_sub}) and (\ref{eq:T_thin}) 
demonstrate. This timescale depends only on the debris particle 
properties --- size and density, being longer for bigger and 
denser particles --- and the outer extent of the disk (for 
$R_{out}/R_{in}\approx 5$ one finds disk lifetime approaching 
a Myr).

In the case of an optically thick disk the lifetime can be 
considerably longer since $t_{thick}$ scales linearly with the 
disk mass, see equation (\ref{eq:T_thick}). Thus, massive disks 
require long time to get exhausted by the PR drag alone. 
For example, an optically thick disk created by the tidal 
disruption of an asteroid with the
diameter of 300 km would have a mass of $4\times 10^{22}$ g 
(assuming bulk density of 3 g cm$^{-3}$). Accretion at the 
steady rate $\dot M_Z=3\times 10^7$ g s$^{-1}$ ($\approx 0.5M_\infty$ 
given by the numerical estimate in equation (\ref{eq:mdot_PR})) would
deplete this disk only within 40 Myr. Such massive disks, however, 
are likely to be affected by the interaction with the gas that
is produced by sublimation of solid debris at the inner edge of 
the disk. They may then evolve in a runaway fashion as described in 
Rafikov (2011b) and get exhausted much earlier.

An interesting finding of this work is the universal appearance 
of the sharp outer edge in optically thick disks, caused 
by the faster inward migration of the disk annuli having smaller 
optical depth. This outer edge steadily moves inwards which, 
combined with the outward expansion of the optically thin 
tail at small radii, leads to the gradual reduction of the 
radial width of the optically thick part of the debris 
disk. Thus, even an extended disk can turn into a narrow ring 
with time as can be easily seen in Figure \ref{fig:tophat}.

Observational evidence for sharp outer edges would strongly support 
the picture of disk evolution due to the PR drag outlined in 
this work. Interestingly, 
Jura \etal (2007b, 2009) have previously claimed the need for 
existence of the {\it outer optically thin} regions in debris disks
around some WDs. However, inferring the sharpness of the outer disk 
edge and the precise value of its optical depth is a nontrivial 
task since any knowledge about the 
radial density distribution in these disks is based on modeling 
their spectral energy distributions. Inverting disk spectrum
to determine the fine details of the radial distribution of 
debris is a rather ambiguous procedure, additionally 
complicated by the possibility of disk warping or
flaring (Jura \etal 2007b, 2009).

\acknowledgements

The financial support for this work is provided by the Sloan 
Foundation and NASA via grant NNX08AH87G. KVB and RRR
thank Princeton University and Lebedev Physical Institute 
for hospitality.


\appendix

\section{Formation of sharp outer edge.}
\label{app:1}

Equation (\ref{eq:2}) is similar to Hopf (or simple wave) equation 
$\partial y/\partial T+y\partial y/\partial x=0$, which is known to 
result in appearance of multivalued solutions usually interpreted 
as evidence for the shock formation. We may expect something similar 
in our case as well because the nonlinearity is clearly present
in equation (\ref{eq:2}). The equation for characteristic, crossing 
the point ($x_1, y \equiv y_1 $), is
\ba 
xy+\ln(1-xy) = x_1 y + \ln(1-x_1 y) + y^2 T. 
\label{eq:char}
\ea
Since the left hand side of (\ref{eq:char}) is negative for all $xy$,
characteristic does not exist at times
\ba 
T > - \frac{x_1 y + \ln (1-x_1 y)}{y^2} = x_1^2 
\varphi (\tau_{\parallel}(x_1)),
~~~~~~~\varphi(u) \equiv \frac{u-1+e^{-u}}{(1-e^{-u})^2}. 
\ea 
From that we can derive an upper limit on the time at which
formation of a sharp edge occurs:
\ba 
T_{break} < \min [x^2 \varphi(\tau_{\parallel}(x))] 
\approx x_0^2/2 \sim T_{thin}\ll T_{thick}.
\ea
Thus, we expect a discontinuity in surface density to form during 
disk evolution and our numerical results in \S\ref{subsect:thick-num}
support this expectation.

\end{document}